%
%
%
%%%%%%%%%%%%%%%%%%%%%%%%%%%%%%%%%%%%%%%%%%%%%%%%%%%%%%%%%%%%%%%%%%%%%%%
%
%
%  Exact Multiplets of Spontaneously Broken Discrete Global Symmetries: 
%
%                   the Example of $N=2$ Susy QCD
%
%                                 by
%
%                    Adel Bilal and Frank Ferrari
%
%
%%%%%%%%%%%%%%%%%%%%%%%%%%%%%%%%%%%%%%%%%%%%%%%%%%%%%%%%%%%%%%%%%%%%%%%

\input phyzzx

%%%%%%%%%%%%%%%%%%%%%%%%%%%%%%%%%%%%%%%%%%%%%%%%%%%%%%%%%%%%%%%%%%%%
% This will make your PHYZZX pagesize wider and longer
% It is OPTIONAL. It redefines the papers macro
%
\catcode`\@=11 % This allows us to modify PLAIN macros.
\def\papersize{\hsize=40pc \vsize=53pc \hoffset=0pc \voffset=1pc
   \advance\hoffset by\HOFFSET \advance\voffset by\VOFFSET
   \pagebottomfiller=0pc
   \skip\footins=\bigskipamount \normalspace }
\catcode`\@=12 % at signs are no longer letters
\papers
%%%%%%%%%%%%%%%%%%%%%%%%%%%%%%%%%%%%%%%%%%%%%%%%%%%%%%%%%%%%%%%%%%%%
\vsize=23.cm
\hsize=15.cm
\tolerance=500000
\overfullrule=0pt

\Pubnum={LPTENS-96/34 \cr
{\tt hep-th@xxx/9606111} \cr
June 1996}

\date={}
\pubtype={}
\titlepage
\title{{\bf Exact Multiplets of Spontaneously Broken \break
Discrete Global Symmetries: the Example of $N=2$ Susy QCD}}

\author{Adel~Bilal}
\andauthor{Frank~Ferrari}
\vskip .5cm
\address{
CNRS - Laboratoire de Physique Th\'eorique de l'\'Ecole
Normale Sup\'erieure
\foot{{\rm unit\'e propre du CNRS, associ\'ee \`a l'\'Ecole Normale
Sup\'erieure et l'Universit\'e Paris-Sud}}    \break
24 rue Lhomond, 75231 Paris Cedex 05, France  \break
{\tt bilal@physique.ens.fr,\   ferrari@physique.ens.fr}
}

\vskip 0.5cm
\abstract{In these notes, we emphasize the r\^ole of
spontaneous broken global discrete symmetries acting on the moduli space of $N=2$ susy
Yang-Mills theories and show how they can be used, together with the BPS condition, as a
spectrum generating symmetry. In particular, in the strong-coupling region, all BPS
states come in multiplets of this broken symmetry. This played a key r\^ole in the
determination of the strong-coupling spectra.}

\vskip 2.cm

\centerline{\it To appear in the Proceedings of the Second International Sakharov Conference,}
\centerline{\it Moscow, May 1996, based on a talk given by A.B. }

\endpage
\pagenumber=1

 \def\PL #1 #2 #3 {Phys.~Lett.~{\bf #1} (#2) #3}
 \def\NP #1 #2 #3 {Nucl.~Phys.~{\bf #1} (#2) #3}
 \def\PR #1 #2 #3 {Phys.~Rev.~{\bf #1} (#2) #3}
 \def\PRL #1 #2 #3 {Phys.~Rev.~Lett.~{\bf #1} (#2) #3}
 \def\MPL #1 #2 #3 {Mod.~Phys. Lett.~{\bf #1} (#2) #3}

\REF\FB{F. Ferrari and A. Bilal, {\it The strong-coupling spectrum of
Seiberg-Witten theory}, to appear in Nucl. Phys. {\bf B469} (1996),
{\tt hep-th/9602082}.}

\REF\BF{A. Bilal and F. Ferrari, {\it Curves of marginal stability, and weak
and strong-coupling BPS spectra in $N=2$ supersymmetric QCD}, \'Ecole Normale
Sup\'erieure preprint LPTENS-96/22, {\tt hep-th/9605101}.}

\REF\SW{N. Seiberg and E. Witten, {\it Electric-magnetic duality, monopole
condensation, and confinement in $N=2$ supersymmetric Yang-Mills theory}, 
\NP B426 1994 19 , {\tt hep-th/9407087},  {\it Monopoles, duality and chiral
symmetry breaking in $N=2$ supersymmetric QCD}, \NP B431 1994 484 ,
{\tt hep-th/9408099}.}

\REF\BIL{A. Bilal, {\it Duality in $N=2$ susy $SU(2)$ Yang-Mills theory: 
A pedagogical introduction to the work of Seiberg and Witten},  
\'Ecole Normale Sup\'erieure preprint LPTENS-95/53, {\tt hep-th/9601007}.}

\REF\ARG{P.C. Argyres, A.E. Faraggi and A.D. Shapere, {\it Curves
of marginal stability in $N=2$ super-QCD}, preprint IASSNS-HEP-94/103, 
UK-HEP/95-07, {\tt hep-th/9505190};\hfill\break
A. Fayyazuddin, {\it Some comments on $N=2$ supersymmetric
Yang-Mills}, \MPL A10 1995 2703 , {\tt hep-th/9504120};\hfill\break
M. Matone, {\it Koebe $1/4$-theorem and inequalities in
$N=2$ super-QCD}, \PR D53 1996 7354 , {\tt hep-th/9506181}.}

\REF\KLMVW{A. Klemm, W. Lerche, P. Mayr, C. Vafa and N. Warner,
{\it Self-dual strings and $N=2$ supersymmetric field theory},
preprint CERN-TH/96-95, HUTP-96/A014, USC-96/008,
{\tt hep-th/9604034}.}

\def\to{\rightarrow}
\def\Z{{\bf Z}}
\def\R{{\bf R}}
\def\C{{\cal C}}
\endpage

\noindent{\bf 1. Introduction}

It has long been known that the spontaneous breakdown of a continuous global symmetry in
quantum field theory leads to certain predictions about the spectrum, namely the
existence of massless Goldstone particles. Here, we will report on another
phenomenon related to the spontaneous breakdown of a discrete global symmetry in theories
having a manifold of physically inequivalent vacua.

It has been observed recently [\FB,\BF] that in $N=2$ supersymmetric $SU(2)$
Yang-Mills theories with and without extra matter hypermultiplets, the BPS states in
the strong-coupling region of moduli space come in {\it multiplets of such a broken
discrete global symmetry}.

In the case of pure $N=2$ susy $SU(2)$ Yang-Mills theory, for example, this discrete
global symmetry is a $\Z_8$. A given vacuum is characterized by a non-vanishing 
expectation value $u=\langle {\rm tr\, } \phi^2 \rangle$, where $\phi$ is the
scalar component of the $N=2$ vector multiplet. Now only a $\Z_4$ subgroup leaves this
$u$ invariant while the remaining elements of $\Z_8$ act as a $\Z_2$ : $u\to -u$ and
hence map a quantum field theory with a given vacuum ($u$) to another quantum field theory
with a different vacuum ($-u$). The important point is that the initial $\Z_8$
symmetry ensures that the theories at $u$ and $-u$ are physically equivalent, and in
particular have the same mass spectrum. This fact, together with the
BPS mass condition, allowed us to show [\FB] that for every state at $u$ of electric
and magnetic quantum numbers $n_e$ and $n_m$ and mass $m$, there exists another state
at $-u$ with different quantum numbers $\tilde n_e, \ \tilde n_m$ but same mass $m$.
This then implies the existence  of this state with quantum numbers 
$\tilde n_e, \ \tilde n_m$ at $u$, but with a different mass $\tilde m$. The two states
$(n_e,n_m)$ with mass $m$ and $(\tilde n_e, \tilde n_m)$ with mass $\tilde m$ at the
{\it same} point $u$ of moduli space hence belong to a multiplet of the broken
symmetry. In the weak-coupling region of moduli space, this chain of arguments
continues and one predicts an infinite tower of states (unless $n_m=0$). In the
strong-coupling region, however, the multiplet is really just the above doublet,
or actually a quartet if one distinguishes particles and antiparticles,
forming a representation of the $\Z_8$. In the cases of $N_f=1,2,3$
additional massless quark hypermultiplets, except for the replacement of this $\Z_8$ by
$\Z_{4(4-N_f)}$, the structure is very similar [\BF]. In the remainder of this short
note, we will try to give a flavour of how these properties come about, concentrating
mainly on the pure $SU(2)$ susy Yang-Mills theory, and then outlining the
generalizations to the cases with hypermultiplets.

\endpage

\noindent{\bf 2. BPS states and curve of marginal stability}

For any background material on the $N=2$ susy $SU(2)$ Yang-Mills theory we refer the
reader to the original papers of Seiberg and Witten [\SW] or to [\BIL]. Here, let us
only mention that the $N=2$ susy algebra admits two types of representations for massive
states, namely
long multiplets containing 16 helicity states, and short ones containing four. 
The masses of the short multiplets obey the so-called BPS condition
$m=\sqrt{2}\vert Z\vert$ where $Z$ is the central charge of the susy algebra. These
states are called BPS states. It was shown in [\SW] that this central charge is given
by
$$
Z=n_e a(u) - n_m a_D(u)\ , \quad m=\sqrt{2}\vert Z\vert
\eqn\i
$$
where $n_e$ and $n_m$ are the integer electric and magnetic charge quantum numbers and
$a(u)$ and $a_D(u)$ can be  explicitly given in terms of hypergeometric
functions (see e.g. [\FB,\BF,\BIL]). We denote a BPS state by $(n_e, n_m)$, so that
e.g. the magnetic monopole is $(0,1)$. The mass being given by the modulus of $Z$, the
triangle inequality, together with the conservation of electric and magnetic charges,
implies that a state with $n_e$ and $n_m$ relatively prime cannot decay into other
states and hence is stable. This argument works as long as $a_D$ is not a real
multiple of $a$. If $a_D(u)/a(u)$ is real, decays are much easier to realise and
otherwise stable BPS states can become unstable. The set \C\ of all $u$ on moduli
space such that $a_D(u)/a(u)\in \R$ is called the curve of marginal stability
[\ARG,\FB]. It is  almost an ellipse, symmetric about the origin $u=0$, and goes
through the singular points of moduli space $u=\pm 1$. As long as two points $u$ and
$u'$ can be joined along a path not crossing \C, any (stable) BPS state $(n_e, n_m)$
existing at $u$ must also exist at $u'$ and vice versa, since one can adiabatically
deform the theory along that path on moduli space and the BPS state will always remain
stable. This obviously is not true if the path has to cross the curve \C. Hence the
curve separates the moduli space into two regions of constant BPS spectra. By the
latter we mean the set of quantum numbers $(n_e, n_m)$ that do exist, not the actual
mass spectrum. We refer to the set of BPS states inside the curve as the
strong-coupling spectrum and to the set outside the curve as the weak-coupling
spectrum. We will show how these spectra are organized into multiplets of the broken
discrete symmetry.

\endpage
\noindent{\bf 3. Broken discrete symmetries}

The origin of the discrete symmetry is a continuous global $U(1)_R$ R-symmetry of the
classical $N=2$ susy Yang-Mills action, under which the scalar $\phi$ has charge two, 
transforming as $\phi\to e^{2i\alpha}\phi$. However, this symmetry does
not survive quantization since the one-loop and instanton contributions are only
invariant under a discrete subset with $\alpha={2\pi\over 8} k,\ k\in \Z$. Thus 
only a $\Z_8$ subgroup remains from the original anomalous $U(1)_R$. This $\Z_8$ is a true
quantum symmetry of the Hamiltonian and of the action. It acts as $\phi^2\to (-)^k
\phi^2$. A given vacuum  has a non-vanishing expectation value of $\phi^2$,
i.e. $u=\langle {\rm tr\, }\phi^2\rangle \ne 0$, and this spontaneously breaks
$\Z_8$ to $\Z_4$, since it is clear that only those elements in $\Z_8$ corresponding
to even $k$ leave the vacuum invariant. The other elements, corresponding to the
quotient $\Z_2$, act as $u\to -u$ and map a given vacuum to another, but physically
equivalent vacuum.

This implies that for any BPS state $(n_e, n_m)$ at $u$, there must be some (other) BPS
state $(\tilde n_e, \tilde n_m)$ at $-u$ with the same mass:
$$
\vert \tilde n_e a(-u) - \tilde n_m a_D(-u) \vert 
= \vert n_e a(u) - n_m a_D(u) \vert \ .
\eqn\ii
$$
Since the mass formula is given in terms of the symplectic invariant of $(n_e, n_m)$
and $(a_D, a)$ it is clear that eq. \ii\ implies the existence of a matrix $G\in Sp(2,\Z)$ 
such that
$$
\pmatrix{ \tilde n_e \cr \tilde n_m \cr}
=\pm G \pmatrix{ n_e\cr n_m\cr} \quad , \quad 
\pmatrix{ a_D\cr a\cr} (-u) = e^{i\omega}\,  G \pmatrix{ a_D\cr a\cr} (u)
\eqn\iii
$$
with $e^{i\omega}$ some phase. From the explicit expressions of $a_D$ and 
$a$ in terms of
hypergeometric functions one finds\foot{
Note that the $\Z_2$ symmetry simply amounts to shifting the $\theta$ angle by $2\pi$.} 
[\FB]
$$
G= G_{W,\epsilon} \equiv \pmatrix{ 1&\epsilon\cr 0& 1\cr}
\eqn\iv
$$
where $\epsilon =\pm 1$ depending on whether $u$ is in the upper or lower half plane,
and the subscript $W$ refers to the weak-coupling region outside the curve \C, since
inside the curve there is a slight subtlety discussed below. According to the argument
just given, the existence of a BPS state $(n_e, n_m)$ at $u$ with mass $m$ implies the
existence of a BPS state $(\tilde n_e, \tilde n_m)= \pm G_{W,\epsilon}\, (n_e, n_m)$ at
$-u$ with the same mass $m$. But if $u$ is outside the curve, then so is $-u$, and we
know that the same state $(\tilde n_e, \tilde n_m)$ must also exist at $u$ although
with a different mass $\tilde m$. Thus the spontaneously broken symmetry $\Z_8$,
through the broken generators leading to the $\Z_2$ acting on the moduli space,
together with the BPS condition, allows us to deduce the existence of $G_{W,\epsilon}\, 
(n_e, n_m)$ from the existence of $(n_e, n_m)$ at the {\it same} point of moduli
space, i.e. for the same quantum field theory with the same vacuum. Given the form of
the matrix $G_{W,\epsilon}$, starting from the monopole $(0,1)$ this leads to a whole
``infinite multiplet" containing all dyons $(n,1)$. In this sense one might call the
$\Z_2$ symmetry realised through the $G_{W,\epsilon}$ matrices a spectrum generating
symmetry. On the other hand, the W-boson $(1,0)$ is a singlet: $G_{W,\epsilon}
(1,0)=(1,0)$.

\vskip 3.mm
\noindent{\bf 4. Strong-coupling multiplets}

Things become even more interesting in the strong-coupling region inside the curve \C,
where other methods like semi-classical quantization do not apply.
This region is separated into an upper and lower one due to the cut of the function
$a(u)$ on the real line. As a consequence, one has to introduce two different
descriptions of the same BPS state. If a state is described by $(n_e, n_m)$ below the
cut, it will be described by $M_1(n_e, n_m)$ above the cut, where $M_1=\pmatrix{\hfill
1&0\cr -2&1\cr}$ is the monodromy matrix around $u=1$. Repeating now the argument, the
existence of $(n_e, n_m)$ at $u$ (with ${\rm Im} u>0$) implies the existence of
$G_{W,\epsilon}\, (n_e, n_m)$ at $-u$. This same BPS state must then also exist at $u$,
but at $u$ it is described as $M_1 G_{W,\epsilon}\, (n_e, n_m)$. Similarly, if one starts
with ${\rm Im} u<0$. Hence the weak-coupling $G$ matrix gets replaced by
$$
G_{S,\epsilon} = (M_1)^\epsilon \, G_{W, \epsilon} = \pmatrix{\hfill 1&\hfill\epsilon\cr
-2\epsilon& -1\cr} \ .
\eqn\v
$$
This matrix has the property that $G_{S,\epsilon}^2 =-{\bf 1}$. As a consequence, the
strong-coupling multiplets are quartets (or doublets if one considers the
hypermultiplets which contain both the particles and their antiparticles), 
and all strong-coupling  hypermultiplets come in pairs. (Of course, there is also the
everywhere massless ``photon" vector multiplet with $n_e=n_m=0$ which is a singlet.)
For $\epsilon=+1$ e.g., these pairs are 
$$\pm \pmatrix{n_e\cr n_m\cr} \quad \longleftrightarrow \quad \pm\,  G_{S,+} 
\pmatrix{n_e\cr n_m\cr} 
=\pm \pmatrix{ n_e+n_m\cr -2n_e-n_m\cr} \ .
\eqn\vi
$$
We insist that for a given point $u$ these two states have different masses. The two
states responsible for the singularities on the moduli space, namely the magnetic
monopole $\pm (0,1)$ and the dyon $\pm (1,-1)$, form such a $\Z_2$ doublet. In [\FB] it was
shown that this is the only doublet in the strong-coupling spectrum, since for any
other doublet \vi\ one or the other partner  becomes massless somewhere on the
curve \C\ of marginal stability
This would lead to extra singularities that we know are not there. Hence these other
doublets cannot exist. This provided a simple determination of the
strong-coupling spectrum [\FB]. Later on, these results were also obtained  
within a string theory context [\KLMVW].

\vskip 3.mm
\noindent{\bf 5. Inclusion of massless quark hypermultiplets}

The previously discussed results have been generalized [\BF] to the inclusion of
$N_f=1,2,3$ massless quark hypermultiplets in the defining representation of the gauge group
$SU(2)$ (asymptotically free theories). In each case, there is a similar curve of
marginal stability separating the Coulomb branch of moduli space into a weak-coupling
region (outside the curve) and a strong-coupling region (inside the curve). The
$U(1)_R$ symmetry is anomalous, and so is the parity operation of the flavour symmetry
$O(2N_f)$. The combined anomaly free subgroup is the discrete $\Z_{4(4-N_f)}$ acting
again non trivially on $\phi$, and the quantum flavour symmetry is 
$Spin(2N_f)$. The vacuum expectation value $u=\langle
{\rm tr\, } \phi^2\rangle$ spontaneously breaks $\Z_{4(4-N_f)}$ to $\Z_4$ with
the quotient $\Z_{4-N_f}$ acting as a symmetry on the Coulomb branch of moduli space,
relating physically equivalent theories. For $N_f=1$ this is a $\Z_3$ acting as $u\to
e^{2\pi i/3} u$, for $N_f=2$ it is a $\Z_2$ acting as $u\to -u$, while there is no
such symmetry for $N_f=3$. As a consequence, the strong-coupling  hypermultiplets 
form triplets
for $N_f=1$ and doublets for $N_f=2$ [\BF]. In these two cases the strong-coupling
spectrum just contains the one triplet or doublet that groups the states responsible
for the singularities. For $N_f=3$ again, the strong-coupling spectrum only contains
the two states responsible for the singularities, but here they do not form any
multiplet.

\vskip 3.mm
\noindent{\bf 6. Conclusion}

We hope to have convinced the reader of the extreme usefulness of the broken
symmetries acting on moduli space, in particular in the strong-coupling regions where
they allow for a simple determination of the spectra of BPS states.

\ack
One of us (A.B.) is grateful to the organizers of the Second International Sakharov
Memorial Conference for the invitation to present the material covered in these notes.

\refout

\end